# The observation of a positive magnetoresistance and close correlation among lattice, spin and charge around $T_C$ in antipervoskite SnCMn$_3$

B. S. Wang,[1] P. Tong,[1,2*] Y. P. Sun,[1*] X. B. Zhu,[1] W. H. Song,[1] Z. R. Yang[1] and J. M. Dai[1]

[1] *Key Laboratory of Materials Physics, Institute of Solid State Physics, and High Magnetic Field Laboratory, Chinese Academy of Sciences, Hefei 230031, People's Republic of China*

[2] *Department of Physics, University of Virginia, Charlottesville, VA 22904, USA*

**Abstract**

The temperature dependences of magnetization, electrical transport, and thermal transport properties of antiperovskite compound SnCMn$_3$ have been investigated systematically. A positive magnetoresistance (~11%) is observed around the ferrimagnetic-paramagnetic transition ($T_C \sim 280$ K) in the field of 50 kOe, which can be attributed to the field-induced magnetic phase transition. The abnormalities of resistivity, Seebeck coefficient, normal Hall effect and thermal conductivity near $T_C$ are suggested to be associated with an abrupt reconstruction of electronic structure. Further, our results indicate an essential interaction among lattice, spin and charge degrees of freedom around $T_C$. Such an interaction among various degrees of freedom associated with sudden phase transition is suggested to be characteristic of Mn-based antiperovskite compounds.

*Corresponding author. Tel: +86-551-559-2757; Fax: +86-551-559-1434.

E-mail address: tongpeng@issp.ac.cn (P. Tong), ypsun@issp.ac.cn (Y. P. Sun)

## 1. Introduction

Recently, the antiperovskite intermetallic compounds $AXM_3$ (A, main group elements; X, carbon, boron or nitrogen; M, transition metal) have attracted considerable attention due to lots of interesting properties, such as superconductivity,[1,2] non-Fermi liquid behavior,[3] strong electron-electron correlation,[4] giant magnetoresistance (MR),[5,6] large negative magnetocaloric effect (MCE),[7-10] giant negative thermal expansion,[11-13] magnetostriction,[14] and nearly zero temperature coefficient of resistivity.[15]

$SnCMn_3$ belongs to the family of Mn-based antiperovskite compounds which have been investigated for several decades. According to the results of neutron diffraction measurement, $SnCMn_3$ has a noncollinear ferrimagnetic (FIM) ground state, being constituted of an antiferromagnetic (AFM) lattice and a ferromagnetic (FM) one with perpendicular to each other.[16] Although equivalent in crystallography, Mn atoms in the unit cell of $SnCMn_3$ can be divided into two types. One type of Mn atoms has a FM arrangement with $0.65\pm0.15\mu_B$/Mn along the [001] direction. The other type has a "square configuration" in the (001) plane with $2.4\mu_B$/Mn and a FM moment of $0.2\mu_B$/Mn along the [001] direction.[16,17] With increasing the temperature, a sudden first-order transition from FIM state to paramagnetic (PM) state occurs at $T_C$, which is concomitant with the shrinkage of unit cell volume.[17] As reported recently by Li et al., partial substitution of Zn for Sn or changing Mn/Sn ratio can significantly change the type of the magnetic transition and the transition temperature of $SnCMn_3$.[6] Most recently, we observed a large MCE (~133mJ/cm$^3$ K in a magnetic field of 48 kOe) near the FIM-PM transition in this system.[18] However, there have been few investigations focusing on the nature of the first-order FIM-PM magnetic transition in parent $SnCMn_3$.

In this paper, we report a systematic study of the magnetization and transport properties of $SnCMn_3$. A positive MR (~11% at 50kOe) is observed near $T_C$ as a consequence of the field-induced phase transition and lattice shrinkage. Meanwhile, an abrupt reconstruction of electronic structure around $T_C$ was confirmed by the analysis of Seebeck and normal Hall coefficients, both of which undergo a sharp change near $T_C$. Furthermore, our experiments suggest a close correlation among lattice, spin, and charge degrees of freedom in the vicinity of $T_C$.

## 2. Experiments

Polycrystalline sample of SnCMn$_3$ was prepared from powders of Sn (4N), graphite (3N) and Mn (4N). The detailed preparation has been reported elsewhere.[18] The magnetic measurements were performed on a Quantum Design superconducting quantum interference device (SQUID) magnetometer (1.8 K ≤ $T$ ≤ 400 K, 0 ≤ $H$ ≤ 50kOe). The resistivity, Seebeck coefficient, thermal conductivity, and Hall effect were measured in a commercial Quantum Design Physical Property Measurement System (PPMS) (1.8K ≤ $T$ ≤ 800K, 0 ≤ $H$ ≤ 90kOe). Unless specially addressed, the temperature-dependent properties in the following discussion were measured in the cooling process.

## 3. Results and discussion

The temperature dependence of field-cooled (FC) and zero-field-cooled (ZFC) magnetizations measured in an applied magnetic field of $H$=100Oe for SnCMn$_3$ is shown in Fig. 1(a). An obvious magnetic transition can be found around 280 K ($T_C$, defined as the temperature where |$dM/dT$| is maximum) where the magnetization decreases abruptly with increasing the temperature, which is due to the FIM-PM transition.[18] Figure 1(b) presents the temperature dependence of $M(T)$ in the vicinity of FIM-PM transition in ZFC process with different magnetic fields. The result indicates that the FIM-PM transition temperature is sensitive to the external field. Namely, $T_C$ increases from 280 K to 282 K with increasing the magnetic fields from 100 Oe to 48 kOe. The isotherm magnetization curve $M(H)$ of SnCMn$_3$ at 5 K is plotted in the inset of Fig. 1(a). It can be seen that the magnetization is almost saturated when the external field $H$ exceeds 10 kOe. The saturated magnetic moment, 0.41$\mu_B$/Mn, is determined by an exploration of the high field $M(H)$ curve to the zero field. This value locates well between those of FM and AFM sites, in accordance with the FIM structure of the ground state. However, the saturated magnetic moment is quite small when it is compared with the localized Mn moment in the Mn-based perovskite oxides, indicating an itinerant character of the carriers in SnCMn$_3$.

Figure 2(a) shows the temperature dependence of resistivity $\rho(T)$ which was measured at $H$=0 Oe and 50kOe in the cooling process. Basically, $\rho(T)$ shows a metallic behavior except for a sudden drop around $T_C$. Compared with the data reported by Li et al,[6] the lower residual resistivity at 5 K [$\rho$(5K)=96 $\mu\Omega$.cm] and the higher residual resistivity ratio [RRR =$\rho$(300K)/ $\rho$(5K) =3.4] may indicate a better quality of our sample. As shown in Fig. 2(b), a positive MR about 11% (defined as MR= $(\rho_H-\rho_0)/\rho_0$) can be observed around $T_C$ in an external magnetic field of 50 kOe. However, when the temperature is much higher or lower than $T_C$, namely in the stable FIM or PM phases, MR is almost zero. This suggests that the MR is extremely relevant to the magnetic phase transition. Also, the positive MR observed here is quite interesting because the application of external field can suppress the randomness of spins around $T_C$, usually leading to a negative MR.[5,6] Considering the remarkable difference in resistivity between below and above $T_C$, the observed MR could be attributed to any shifts of $T_C$ by external fields. As shown in inset of Fig. 2(a), around $T_C$, the resistivity in PM phase is much lower than in FIM phase. The application of external field can shift $T_C$ to a higher value as evidenced in Fig. 1(b). Therefore, the transition from the low-resistivity state to the high-resistivity one near $T_C$ give rise to the positive MR. At other temperatures far away from $T_C$, the system is in the stable PM or FIM state. Thus, the magnetic filed, even as high as 50 kOe, is not able to produce an obvious MR effect. This interpretation is further confirmed by Fig. 3, in which the magnetic field dependence of the MR (%) at selected temperatures around $T_C$ is presented. It can be seen that the value of MR is very small when $T<$ 279K or $T >$282K since the field-induced magnetic phase transition doesn't happen in the stable FIM or PM phase. However, near $T_C$ (such as at 280K and 281K), the MR increases swiftly with the field $H$ to a certain value. As increasing $H$ further, MR becomes nearly saturated as the field-induced phase transition is completed.

Figure 4(a) shows the normal Hall coefficient $R_H$ as a function of temperature which is measured in an external magnetic field of 10 kOe. It is clear that $R_H$ is negative in the whole temperature range studied, suggesting that the dominant carriers of $SnCMn_3$ are electron-type. Below 100 K, the magnitude of $R_H(T)$ is almost a constant, indicating a normal metallic behavior. As temperature increases, $R_H(T)$ becomes more dependent on the temperature. At first, a sharp change of $R_H(T)$ occurs around $T_C$ (shown in Fig. 4(a)). The strong dependence of $R_H(T)$ on the temperature implies a complex underlying electronic structure with more than one band near Fermi surface containing different types of carriers.[20-22] Regardless of this, the Hall carrier density $n_H$ can be reasonably estimated using the formula $n_H = 1/|eR_H|$,[23] where $e$ is elementary electric charge. Around $T_C$, we find that $n_H$ in high-temperature PM phase is increased by three times in comparison with which in low-temperature FIM phase.[18] Therefore, the ordering of spins leads to, at least partially, the localization of itinerant charges, suggesting a correlation between spin and charge in SnCMn$_3$. This behavior of $R_H(T)$ is analogous to what happened in GaCMn$_3$, where $1/R_H$ of high-temperature FM phase is five times larger than that of low-temperature AFM phase.[5] The temperature dependence of Seebeck coefficient $\alpha(T)$ is shown in Fig. 4(b). Generally, the sign of $\alpha(T)$ can reflect the type of carriers reasonably.[24] At low temperatures, the negative value of $\alpha(T)$ in SnCMn$_3$ indicates that the electron-type carriers are dominant, which is consistent with the result of normal Hall effect. However, at elevated temperatures, $\alpha(T)$ becomes no longer linear temperature-dependent, and shows a broaden minimum near 150 K. The phonon-drag effect, which is usually important at temperatures lower than 100 K,[25] is insufficient to account for the observed broaden minimum in present compound. Alternatively, the existence of a pseudo-gap in the electronic density of state (DOS) might be responsible for such a broaden minimum as observed in Heusler alloys (e.g., Ni$_2$MnGa).[26] In contrast, there is no observable abnormality in $\rho(T)$ curve and in the lattice constant around 150 K.[6,18] With further increasing the temperature, $\alpha(T)$ jumps suddenly at $T_C$ from a negative value in FIM state to a positive one in PM state. This positive value of $\alpha(T)$ above $T_C$ is sharply inconsistent with the negative normal Hall coefficient. The different signs of Seebeck and normal Hall coefficients may be a result of multiband in electronic structure. Indeed, previous theoretical study have revealed that the Fermi surface of SnCMn$_3$ consists of two electron surfaces and one hole surface.[19] It is of particular complex to interpret quantitatively the difference between the signs of the Seebeck and normal Hall coefficients exhibit in the framework of multiband. In order to simplify the problem, we assume the electrons in the two electron bands are the same in nature. Thus the two-band model, i.e., two-carrier model, is employed in the following discussion to generate a qualitative comprehension of the opposite signs for the normal Hall and Seebeck coefficients.

In the two-band framework, the normal Hall coefficient $R_H$, Seebeck coefficient $\alpha$, and electronic conductivity $\sigma$, can be written as follows,[21, 22, 27]

$$R_H = \frac{\mu_+^2 n_+ - \mu_-^2 n_-}{e(\mu_+ n_+ + \mu_- n_-)^2}, \quad (1)$$

$$\alpha = \frac{\alpha_+ \mu_+ n_+ + \alpha_- \mu_- n_-}{e(\mu_+ n_+ + \mu_- n_-)}, \quad (2)$$

$$\sigma = \sigma_+ + \sigma_- = e(\mu_+ n_+ + \mu_- n_-), \quad (3)$$

Where $n$, $\mu$, $\alpha$, and $\sigma$ represent for the concentration, mobility, Seebeck coefficient, and electrical conductivity for electron (-) or hole (+) bands, respectively. In the simplest case for free electrons, $\alpha_\pm = \pm \pi^2 \kappa_B^2 T / 3eE_F$.[28] After the FIM-PM transition, $R_H$ is negative, but $\alpha$ is positive. Thus, the equations (1) and (2), require that $\mu_+^2 n_+ < \mu_-^2 n_-$ and $\mu_+ n_+ > \mu_- n_-$, respectively. Combining these two relationships, one can get $n_+ > n_-$ and $\mu_+ < \mu_-$. While below $T_C$, the electron-type carriers are dominant as manifested above. This indicates the concentration of hole-type carrier increases suddenly with a reduced mobility when the system passes the FIM-PM transition. This deduction can account for the corresponding sudden increase of $\sigma$ in PM state, which gives rise to the decrease of absolute $R_H$ value in PM state on the base of equation (1). Practically, however, the Seebeck coefficient is strongly dependent on the energy. Provided that the scattering time is independent of energy, Seebeck coefficient for each band can be written as,[28]

$$\alpha_\pm = \pm \frac{\pi^2 \kappa_B^2 T}{3e} \left(\frac{\partial \ln N_\pm(E)}{\partial E}\right)_{E_F}, \quad (4)$$

where $N(E)$ is the DOS for electron (-) or hole (+) bands. $k_B$, and $E_F$ is the Boltzmann constant, and Fermi energy, respectively. The reduced mobility (i.e., enhanced effective mass) of hole-type carriers after FIM-PM transition is just a reflection of swift change in DOS with energy at $E_F$.[29] Such a swift variation of DOS with energy could increase the absolute value of $\alpha_+$, thus contribute more or less to the observed positive Seebeck coefficient above $T_C$. It is worthy to note that the changes of carrier density can also be regarded as a main result of modification of electronic structure. As a result, the abnormalities of transport behavior at $T_C$ can be tentatively explained in terms of the two-band model. Nevertheless, a thorough understanding of these abnormal behaviors in normal Hall and Seebeck coefficients lies on further theoretical calculation and special experiments, e.g., De-Haas-van Alphen resonance and angle-resolved photoemission spectroscopy which refer directly to the detailed electronic structure. In spite of this, our result reveals manifestly an abrupt reconstruction of electronic structure, in the vicinity of $T_C$.

Figure 5 displays the temperature dependent thermal conductivity $k(T)$. Observably, a broaden peak and a kink exist around 50 K and $T_C$, respectively. In general, the total thermal conductivity for a metal can be expressed as a sum of lattice ($\kappa_L$) and electronic ($\kappa_e$) terms, i.e., $\kappa(T)$

= $\kappa_L(T) + \kappa_e(T)$. The electronic thermal conductivity $\kappa_e$ can be estimated using the Wiedemann-Franz law $\kappa_e \rho / T = L_0$ (where $L_0 = 2.45 \times 10^{-8} W\Omega K^{-2}$ is the Lorentz number) and resistivity $\rho(T)$. The lattice thermal conductivity $\kappa_L$ is taken as the difference between $\kappa$ and $\kappa_e$. It is apparent that in the whole temperature range $\kappa(T)$ originates mainly from the $\kappa_L$. So does the broadened peak around 50K. This peak is a typical feature for the reduction of thermal scattering in solids at low temperatures. The kink at $T_C$ can be mainly ascribed to the contribution from $\kappa_L$. While $\kappa_e$ goes through an enhancement at $T_C$ as temperature increases, indicative of an increase in carrier density, which is in good agreement with the behavior of normal Hall carrier density. This behavior implies the interaction between lattice (phonons) and conduction charges in the vicinity of $T_C$, which could rise from phonon softening, leading to a Kohn anomaly near the phase transition.[26, 30]

As discussed above, the analysis of magnetic, electrical transport and thermal transport properties of SnCMn$_3$ indicates that the lattice (phonon), spin and charge interact with each other in the vicinity of $T_C$. Such an interaction between various degrees of freedom would reach a state of equilibrium, which can be manipulated by external disturbances such as magnetic field, pressure and heat etc. As an example, the MR effect discussed above can be thought as a result of field-induced phase transition. Analogously, the prototype Mn-based antiperovskite compound GaCMn$_3$ is considered to be unique due to its first-order AFM-FIM phase transition, at which lattice, spin and charge are correlated with each other.[31] As reported previously, the sudden magnetic/structural phase transitions are universally observed in Mn-based antiperovskite compounds.[11-16] Along with our research on SnCMn$_3$, the interaction among spin, lattice and charge seems to be universal in this type of intermetallic compounds. In this sense, more researches on Mn-based antiperovskite compounds are desirable for a better understanding of the correlations among various degrees of freedom, as well as for exploring new materials with interesting properties, e.g., GMR, MCE.

## 4. Conclusions

In summary, the magnetic, electrical and thermal transport properties of antiperovskite compound SnCMn$_3$ have been investigated. The swift reconstruction of electronic structure has been identified in the vicinity of the first-order FIM-PM transition ($T_C \sim$ 280 K), which can account for the abnormal behaviors in transport properties in terms of the two-band model. Further, our result clearly indicates a close correlation among lattice, spin and charge around $T_C$. The positive MR of ~11% at 50 kOe observed near $T_C$ can be attributed to the field-induced magnetic phase transition. The essential correlations among various degrees of freedom with respect to sharp phase transition are thought to be inherent in Mn-based antiperovskite compounds, which thus are desirable for further studies.

**Acknowledgements**

This work was supported by the National Key Basic Research under contract No. 2007CB925002, and the National Nature Science Foundation of China under contract No.50701042, No.10774146, No.10774147 and Director's Fund of Hefei Institutes of Physical Science, Chinese Academy of Sciences.

**Figure captions**

**FIG. 1.** (Color online) (a) The temperature dependence of magnetization $M$(T) under ZFC and FC processes at $H$ = 100Oe. Inset of (a): the isotherm magnetization curve $M(H)$ at 5K. (b) $M$(T) at different fields around $T_C$ under ZFC process. Arrow indicates the increase of external field applied.

**FIG. 2.** (Color online) (a) The temperature dependence of resistivity $\rho(T)$ measured at $H$=0 Oe and 50kOe in the cooling process. The inset shows the enlargement of resistivity in the vicinity of $T_C$. (b) The magnetoresistance deduced from the result of (a).

**FIG. 3.** (Color online) The MR as a function of the external magnetic field up to 50kOe at 278K, 279K, 280K, 281K, 282K.

**FIG. 4.** (Color online) (a) The normal Hall coefficient as a function of temperature (b) The temperature dependence of Seebeck coefficient $\alpha(T)$ at zero field.

**FIG. 5.** (Color online) The thermal conductivity $\kappa$ as a function of temperature. The lattice ($\kappa_L$) and the electronic ($\kappa_e$, enlarged by 2.5 times) contributions are also shown.

**Figure:**

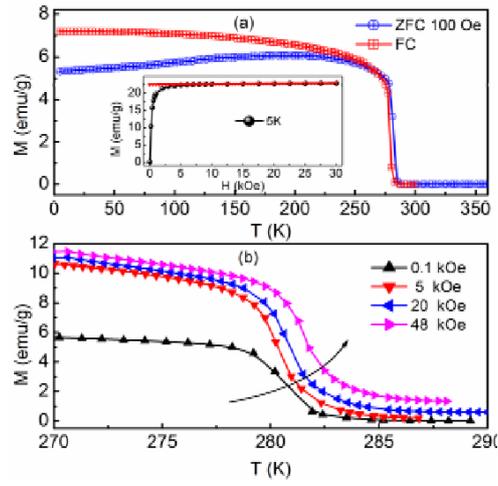

**Figure 1.** B. S. Wang et al.,

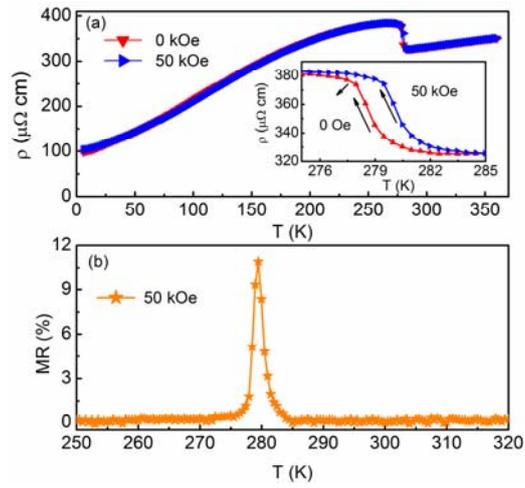

**Figure 2.** B. S. Wang et al.,

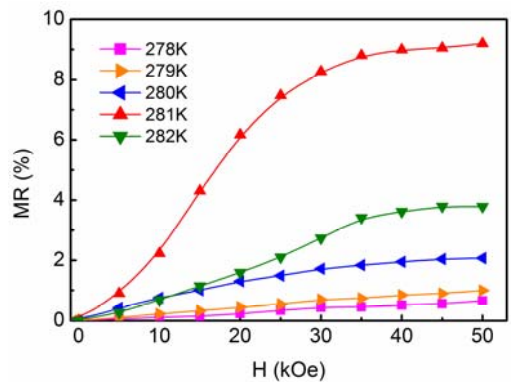

**Figure 3.** B. S. Wang et al.,

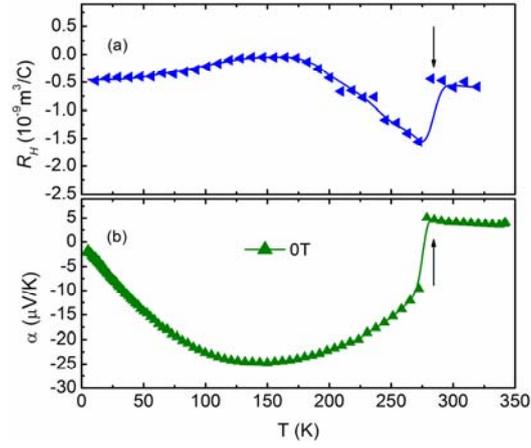

**Figure 4.** B. S. Wang et al.,

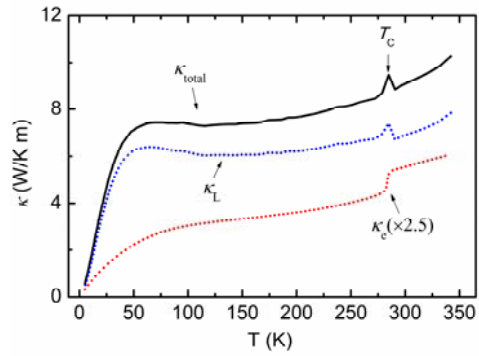

**Figure 5.** B. S. Wang et al.,